\documentclass[]{spie}  

\usepackage{amsmath,amsfonts,amssymb}
\usepackage[colorlinks=true, allcolors=blue]{hyperref}
\usepackage[final]{graphicx}
\usepackage{float}
\usepackage[caption = false]{subfig}
\usepackage{graphicx}
\usepackage[utf8]{inputenc}
\usepackage[french]{babel}
\usepackage{subfig}
\usepackage{txfonts}

\usepackage[section]{placeins}

\newcommand\be{\begin{equation}}
\newcommand\en{\end{equation}}

\newcommand{\arcs}{\hbox{$^{\prime\prime}$}}

\newcommand{\arcm}{\mbox{$^{\prime}$}}

\newcommand{\ks}{\mbox{$K_{\rm s}$}}

\title{Correction of distortion for optimal image stacking in Wide Field Adaptive Optics: Application to GeMS data.}

\author[a]{Anaïs Bernard}
\author[b]{Laurent M. Mugnier}
\author[a]{Benoit Neichel}
\author[a,b]{Thierry Fusco}
\author[a]{Sophie Bounissou}
\author[a]{Manash Samal}
\author[d]{Morten Andersen}
\author[a]{Annie Zavagno}
\author[e]{Henri Plana}

\affil[a]{Aix Marseille Université, CNRS, LAM (Laboratoire d'Astrophysique de Marseille) UMR 7326, 13388, Marseille, France}
\affil[b]{ONERA (Office National d’Etudes et de Recherches Aérospatiales),B.P.72, F-92322 Chatillon, France}
\affil[d]{Gemini Observatory, c/o AURA, Casilla 603, La Serena, Chile}
\affil[e]{Laboratorio de Astrof\'isica Te\'orica e Observacional, Universidade Estadual de Santa Cruz, Rodovia Jorge Amado km16 45662-900 Ilh\'eus BA - Brazil}

\authorinfo{Further author information: (Send correspondence to Anaïs Bernard)\\Anais Bernard: E-mail: anais.bernard@lam.fr\\   Benoit Neichel: E-mail: benoit.neichel@lam.fr}

\begin{document} 

\maketitle
\begin{abstract}
The advent of Wide Field Adaptive Optics (WFAO) systems marks the beginning of a new era in high spatial resolution imaging. The newly commissioned Gemini South Multi-Conjugate Adaptive Optics System (GeMS) combined with the infrared camera Gemini South Adaptive Optics Imager (GSAOI), delivers quasi diffraction-limited images over a field of $\sim$ 2 arc-minutes across. However, despite this excellent performance, some variable residues still limit the quality of the analyses. In particular, distortions severely affect GSAOI and become a critical issue for high-precision astrometry and photometry.  In this paper, we investigate an optimal way to correct for the distortion following an inverse problem approach. Formalism as well as applications on GeMS data are presented.

\end{abstract}

\keywords{Wide Field Adaptive Optics, data post-processing, distortions correction, star forming region, photometry, astrometry, image registration, inverse problems}

\section{INTRODUCTION}  
By using multiple Laser Guide Stars (LGS), Wide Field Adaptive Optics (WFAO), improves the performance of high spatial resolution imaging: The AO-corrected images Field of View (FoV) is increased, as well as the fraction of the sky that can benefit from such correction. The Gemini Multi-Conjugated Adaptive Optics (MCAO) instrument GeMS is the first multi-Laser Guide Star operational system on sky. It has been implemented on the Gemini South telescope and commissionned in 2013. It works with two deformable mirrors conjugated at 0 and 9 km and a sodium-based LGS constellation composed of five spots: four are located at corners of a 60 arcsec square, with the fifth positioned in the center. GeMS, as a facility instrument, can direct its light output to different science instruments installed at the Cassegrain focus of the Gemini South telescope. Combined with the Gemini South Adaptive Optics Imager (GSAOI), it delivers near-diffraction limited images at Near-Infrared (NIR) wavelength (from 0.9 to 2.4 $\mu$m) over a FoV of 85$\arcs \times 85\arcs$. More details about the GeMS/GSAOI system and its commissioning results are described in detail in previous papers (see McGregor et al. (2004)\cite{McGregor2004}, Carrasco et al.(2011)\cite{Carrasco2011a}, D'Orgeville et al. (2012)\cite{DOrgeville2012}, Neichel et al. (2013 \cite{Neichel2013a} and 2014\cite{Neichel2014a}) and Rigaut et al. (2014)\cite{Rigaut2014}).
However, despite the excellent performance of the GeMS/GSAOI system, the correction provided is not perfectly uniform and may generate variable Point Spread Function (PSF) over the field. For instance, we observe in the data an average Full Width Half Maximum (FWHM) over the field ranged between 80 mas and and 145 mas depending on the filter and the natural seeing. The standard deviation associated is typically around 15 mas, corresponding to a variation of 10-15\%. The average Strehl Ratio (SR) ranges between 3\% and 14\%  with a standard deviation from 1\% to 3 \% (see Tab.\ref{tab1}).
Those spatial variations of the PSF are present: 
\begin{enumerate}
\item on single frames, mostly due to residuals from the AO correction ;
\item on stacked images, critically amplified by optical distortions generated by the system ;
\end{enumerate}
Indeed, the optical components present in the instrument and the telescope as well as the GSAOI camera, introduce static and dynamical distortions. They depend on environmental parameters like the LGS spot size or the telescope pointing and may vary from one frame to another. These distortions degrade the resolution and strongly reduce the sensitivity when combining multiple frames. From there, the ability to deal with the spatial variation of the PSF is critical for high-precision astrometry and photometry studies (see Massari et al. (2015)\cite{Massari2015}, Turri et al. (2015)\cite{Turri2015}). Therefore, dedicated tools are needed to properly correct for distortion and preserve as much as possible the initial quality of the data when stacking them.

In a first part of this paper we present a set of data obtained with the GeMS/GSAOI system: a recent observations of a very active and young star-forming region N159W located in the Large Magellanic Cloud (LMC) and that was previously studied by Deharveng et al. (1992)\cite{Deharveng1992}, Testor et al. (2007)\cite{Testor2006}, Chen et al. (2010)\cite{Chen2010a}. 
We obtained deep $J$, $H$, and \ks images in order to study the properties of the cluster stellar members and bring new elements to our understanding of the massive star formation process (for complete study see Bernard et al. (2016) \cite{Bernard2016}).
The N159W field provides a large number of isolated stars and is therefore an ideal case to evidence limitations and experiment new methods of correction of distortions. 
In a second part of this paper we investigate an optimal way to correct for the distortion following an inverse problem approach. The method is based on the work presented in Gratadour et al. ($A\&A$ 2005)\cite{Gratadour2005} on image re-centering, but generalized to all distortion modes. We present here the formalism and simulation results as well as first application on the N159W GeMS/GSAOI data.

\section{OBSERVATIONS AND DATA REDUCTION}
\label{sec:sec1}

\subsection{Observations}

The data used for the analysis were obtained during the night of December 8th 2014 as part of program GS-2014B-C-2 (P.I. B. Neichel). Images corrected for atmospheric turbulence were obtained using the GeMS/GSAOI instrument. Details of the observations are summarized in Table~\ref{tab1}. Images were recorded through $J$, $H,$ and \ks filters.  Each observation consists of one science fields dithered randomly by a 5" $rms$ shift to remove gaps between the detectors and a similar scheme for the sky frames taken 2.5\arcm  away from the science field. The averaged resolution over the field, measured as the FWHM of the stars on single-exposure frames is reported in Table~\ref{tab1} as well as the natural seeing (measured at zenith). The averaged FWHM in \ks  is 90 mas, while the averaged Strehl ratio (SR) is 14\%. Those performance are slightly worst than those expected for the good seeing conditions, but there are consistent with observations at low elevation, as it is the case for the LMC. The coordinates of the center of the field are RA = 05$^h$39$^m$40$^s$ ; DEC = -69\degre 45\arcm 55\arcs. We have adjusted the field in order to encompass what was previously observed by Testor et al. (2006)\cite{Testor2006} who previously studied this region. For full study of the stellar content of the region see Bernard et al. (2016) \cite{Bernard2016}.

\begin{table*}[h!]
\caption{Observation details. FWHMs, and SRs averaged over $\sim$ 200 stars uniformly distributed over the field and for all the individual frames. The corresponding standard deviations are also given.}      
\vspace*{0.25cm}   
\label{tab1} 
\begin{tabular}{lrrrrrrrrr}
\hline
\hline
Date &Filter &Individual & Number of & $<$FWHM$>$ & $\sigma_{\rm{FWHM}}$ & $<$SR$>$ & $\sigma_{\rm{SR}}$ & Natural seeing (") & PA \\
     &       & exposure time & frames  &  (mas)  & (mas) & (\%) & (\%) & (\arcs @ 0.55$\mu$m) & (degree) \\
\hline
Dec  8$^{th}$& $J$  & 80s & 17  & 145  & 15 & 3 & 1 & 0.55 & 320\\
      2014   & $H$  & 80s & 14 &  100  & 20 & 9 & 2 & 0.60 & 320\\
                   & \ks & 80s & 17  & 90  & 15 & 14 & 3 & 0.55 & 320 \\                 
\hline
\end{tabular}
\end{table*}

\subsection{Data reduction}

\begin{figure*}[h!]
\begin{center}
\includegraphics[scale=0.07, angle=-45]{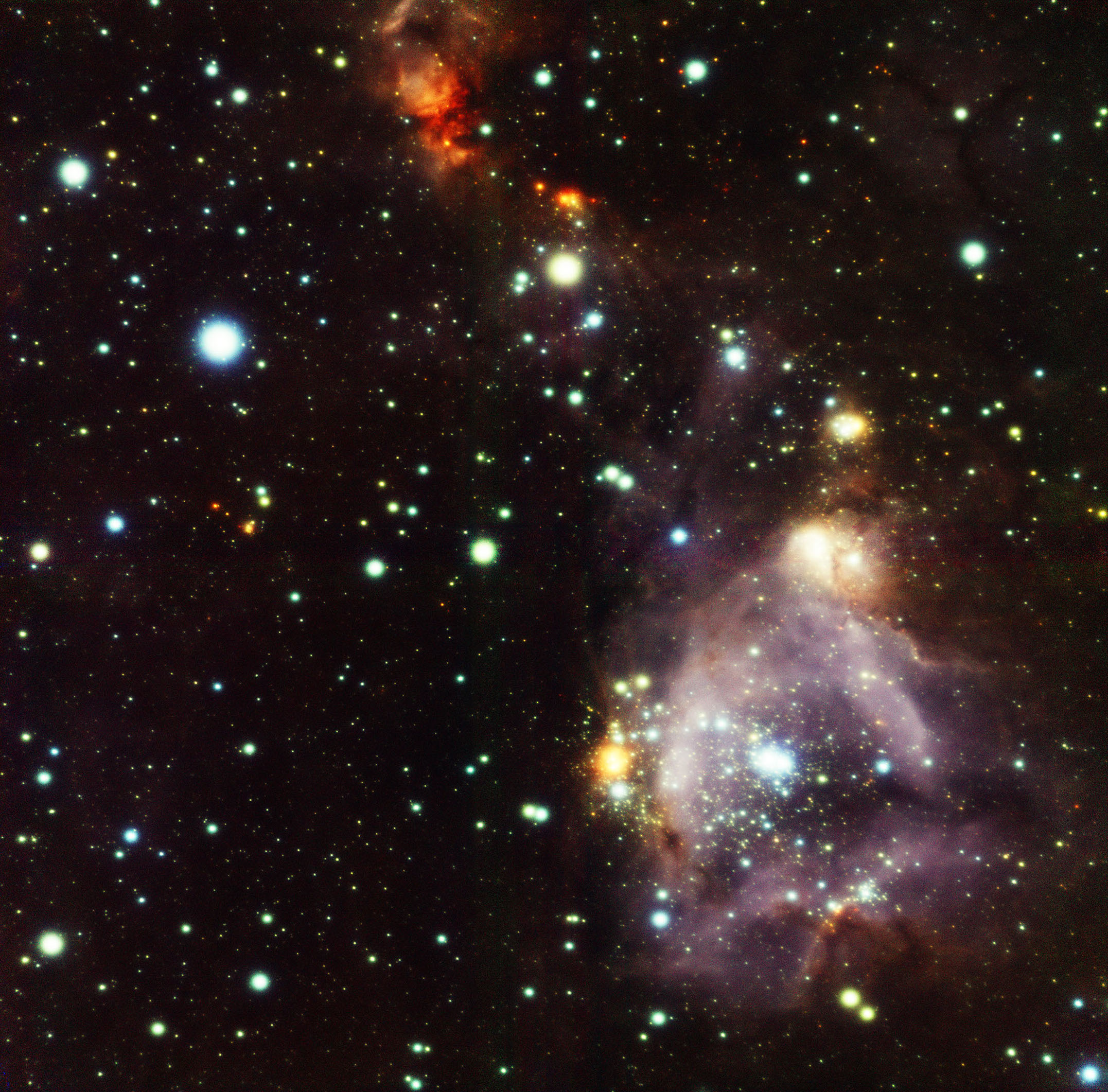}
\vspace*{0.3cm}
\caption{N159W false three-color GeMS/GSAOI image combining $J$ (blue), $H$ (green), and \ks (red). North is up and east is left. The width of the image (East to West) is 90 arcs. }
\label{fig:threecolor}%
\end{center}
\end{figure*}

Data reduction was done using home-made procedures developed in \texttt{Yorick} (Munro \& Dubois 1995) \cite{Munro1995}. The different steps are:
(i) creation of a master flat image based on sky flat images taken durind twilight of the same night;
(ii) creation of a master sky frame based on the dedicated sky images;
(iii) correction of the science frame from the master flat and the master sky, as well as for detector non-linearities and different gains between each detector. 
(iv) application of an instrumental distortion correction. This last step was done according to the method described in Neichel et al. (2015) \cite{Neichel2015a}. 
Following this procedure, each individual image has been reduced, and eventually combined by filter, to produce three long-exposure and three short-exposure reduced images. Figure \ref{fig:threecolor} shows the final three-color image built from the GeMS/GSAOI data in $J$ (blue), $H$ (green), and $Ks$ (red) bands.

The method of distortion correction presented in Neichel et al. (2015) \cite{Neichel2015a} show good results but
is not precise enough for high-precision astrometry and photometry studies, especially on sparse fields. Indeed,  the main limitation of this method is that it requires a large number of polynomial orders (15 degrees of freedom) to converge toward a minimal residual distortion (see Neichel et al. (2014b) \cite{Neichel2014b}). This means that a large number of well isolated and bright stars are needed, which is not often the case. Another inconvenient lies in the fact that the correction is based on a reference chosen within the data and so, not free from distortions. 
In the next section, we present a new method of distortion correction currently in development
 that is addressing both limitations.

\section{TOWARDS AN OPTIMAL DISTORTION CORRECTION} 
\label{sec:sec3}

\subsection{Distortion characterization}
Distortions in GeMS/GSAOI data have been estimated to be on the order of 200 mas (peak value) and are coming from multiple sources. First, the off-axis parabola present in the AO bench as well as the deformable mirror conjugated in altitude are introducing low spatial orders of distortions. Then, from the analysis carried in Neichel et al. (2014b) \cite{Neichel2014b}, it has been showed that higher spatial orders are also present (around 15 degrees of freedom). Part of these high order distortions might be due to the infrared camera and errors in the gaps correction. Indeed, the camera is composed of four distinct detectors separated by gaps. During the data reduction process, the four detectors are considered to be perfectly positioned, which might not be the case in reality. During the dithering of the telescope, a star moving from one detector to another between the different frames might introduce high orders of distortion. Finally, another source of errors comes from dynamical distortions, which depends on the Natural Guide Stars (NGS) constellation and environmental factors like the telescope pointing and the dithering. Therefore, the correction of distortion must take into account a large number of degree of freedom and be specific to each epoch. 

\vspace{0.5cm}


\subsection{Description of the model}

Aiming to increase the photometric and astrometric performance during the scientific analysis of GeMS/GSAOI data, we now investigate an optimal way to correct for the distortions following an inverse problem approach based on the work developed by D. Gratadour  and L. Mugnier. The original work of Gratadour et al. (2005) \cite{Gratadour2005}, was focusing on image re centering while here we  generalize it to all distortion modes. The model is based on polynomial distortions, as follows : 

\begin{equation}
\label{eq:dist}
X_{i,j}^{data}=\begin{pmatrix}x^{data}_{i,j}\\y^{data}_{i, j} \end{pmatrix}=\begin{pmatrix}\sum\limits_{n=0}^{N}  \sum\limits_{k=0}^{n} a_{i,k,n,x}~{(x_{j}^{ref})}^{k}{(y_{j}^{ref})}^{n-k}\\\sum\limits_{n=0}^{N}  \sum\limits_{k=0}^{n} a_{i,k,n,y}~{(x_{j}^{ref})}^{k}{(y_{j}^{ref})}^{n-k} \end{pmatrix}+noise
\end{equation}

With : \vspace{0.3cm}
\\ {$X_{i,j}^{data}=\begin{pmatrix}x^{data}_{i,j}\\y^{data}_{i, j} \end{pmatrix}$, the coordinates of star $j$ in image $i$;\vspace{0.1cm}
\\{$x_{j}^{ref}$ and $y_{j}^{ref}$ the $x$ and $y$ distortion-free coordinates of star $j$. Both coordinates will be mentioned as $X^{ref}_{j}$ hereafter;} \vspace{0.1cm}
\\{$a_{i,k,n,x/y}$, the distortion coefficients on axis $x/y$ of image $i$ for the $(k,n)$ distortion mode. The whole set of distortion coefficients associated to image $i$ will be mentioned as $A_i=\{a_{i, k,n, x}, a_{i, k,n,y}\}_{\substack{0\leq n \\ 0 \leq k}}$ hereafter;} \vspace{0.1cm}
\\{$N$, the highest polynomial order. $N=0$ corresponds to re centering.} \vspace{0.1cm}
\vspace{0.5cm}

\noindent Assuming that the noise on the measurements of $X_{i,j}^{data}$ is white and Gaussian (which is reasonable because the positions measurements combine many pixels), the Maximum Likelihood estimation of the unkows boils down to the minimization of a criterion defined as follow: 

\begin{equation}
J\left((X^{ref}_j)_{\substack{1\leq j \leq Nstar}},(A_{i})_{\substack{1\leq i \leq Nim}}\right)=\left\{
    \begin{array}{ll}
        \sum\limits_{i=1}^{N_{im}}  \sum\limits_{j=1}^{N_{star}} w_{i,j}\left|x_{i,j}^{data}-\sum\limits_{n=0}^{N}  \sum\limits_{k=0}^{n} a_{i,kn,x}~{(x_{j}^{ref})}^{k}{(y_{j}^{ref})}^{n-k}\right|^2 \\
        \sum\limits_{i=1}^{N_{im}}  \sum\limits_{j=1}^{N_{star}} w_{i,j}\left|y_{i,j}^{data}-\sum\limits_{n=0}^{N}  \sum\limits_{k=0}^{n} a_{i,kn,y}~{(x_{j}^{ref})}^{k}{(y_{j}^{ref})}^{n-k}\right|^2 \\
    \end{array}
\right.
\end{equation}

With : \vspace{0.3cm}
\\{$Nim$, the number of images;} \vspace{0.1cm}
\\{$Nstar$, the number of stars;} \vspace{0.1cm}
\\{$w_{i,j}, $ the weighting coefficient of star $j$ in image $i$;} \vspace{0.1cm}
\\~

\noindent The method allow the joint estimation of both:
\begin{enumerate}
\item the distortion coefficients associated to each frame $(A_i)_{\substack{1\leq i \leq Nim}}$;
\item the distortion-free positions of each star in the field $(X^{ref}_j)_{\substack{1\leq j \leq Nstar}}$;
\end{enumerate}
The number of degrees of freedom is given by $d=(N+1)(N+2)$.

\subsection{Simulation results}
\label{sec:simu}
The algorithm is built as an alternate minimization on: (i) each image $i$ independently, to estimate its distortion coefficients $A_i$; (ii) each star $j$ independently, to estimate its distortion-free position $X^{ref}_j$.
In this section, we study the behavior of the algorithm with respect to noise. To do so, we implemented the method considering $d=20$ degrees of freedom (i.e., $N=3$), according to results published in previous studies (see Neichel et al. (2014)\cite{Neichel2014b}) and we worked on simulated data. We evaluate here :  
\begin{enumerate}
\item This method's ability to estimate the distortion parameters i.e., the distortion coefficient, $(A_i)_{\substack{1\leq i \leq Nim}}$ and the distortion-free position of the stars, $(X^{ref}_j)_{\substack{1\leq j \leq Nstar}}$; 
\item The quality of the distortion correction provided by this method;
\end{enumerate}

\subsubsection{Estimation of the distortion}
\label{sec:estimation}
To evaluate the method's ability to estimate the distortion coefficients, $(A_i)_{\substack{1\leq i \leq Nim}}$ and the distortion-free positions of the stars, $(X^{ref}_j)_{\substack{1\leq j \leq Nstar}}$, the idea is to introduce a well-known distortion on one fake reference image as describe hereafter.  We first built one simulated image composed of noise-free PSFs randomly distributed over the field. We used this image as reference where the position of the PSFs are the distortion-free positions of the stars $(X^{ref}_j)_{\substack{1\leq j \leq Nstar}}$. We then introduced random distortion coefficients $A_i$ on this image using Equation \ref{eq:dist} to build the $X_{i,j}^{data}$. The applied distortion is lower than 10 pixels ($\sim$ 200 mas). 
The process is repeated several times to obtain a cube of distored images where the PSFs positions are given by $(X^{data}_{i,j})_{\substack{1\leq i \leq Nim\\1\leq j \leq Nstar}}$. Using these star positions as input and after running the method, we found that both distortion parameters, the distortion coefficients of each frames, $(A_i)_{\substack{1\leq i \leq Nim}}$ and the distortion-free position of each PSFs, $(X^{ref}_j)_{\substack{1\leq j \leq Nstar}}$, are estimated with the numerical precision of $10^{-7}$ pixel. 

In order to predict the behavior of the algorithm confronted to real data, we now build noisy measurements by adding a Gaussian noise on the measured positions of the PSFs $(X^{data}_{i,j})_{\substack{1\leq i \leq Nim\\1\leq j \leq Nstar}}$ previously used. We do this for varying levels of noise ($\sigma_{noise}$=[0,0.05,0.1,0.2,0.5,1,2] pixels).
Once the noise is added, the distortion coefficient, $(A_i)_{\substack{1\leq i \leq Nim}}$ and the distortion-free position of the PSFs, $(X^{ref}_j)_{\substack{1\leq j \leq Nstar}}$ are estimated following the same method as described above. We then repeat the process 100 times for one given $\sigma_{noise}$ and derive an average error of estimation associated to the given $\sigma_{noise}$.  

Fig.\ref{fig:noiseprop} shows the error of estimation versus the noise for simulations done with $N=3$ (where $N$ is the highest polynomial order), $Nim=10$ and $Nstar=20$. Fig.\ref{fig:noiseprop} (a) shows the error on the first mode of distortion estimation: the shift (the error is averaged between the shift on the $x$ axis and the shift on the $y$ axis). 
Fig.\ref{fig:noiseprop} (b) shows the error on the distortion-free position of the PSFs estimation. 
In both cases, the error represents the distance, in pixels, between the true value and the estimated value. It is quadratically averaged on the images and on the PSFs (respectively for the distortion coefficients and the distortion-free positions of the PSFs) and on the 100 noise values. The errors are plotted versus the standard deviation of the added noise.

\begin{figure}[h!]
\begin{center}
\subfloat[]{\includegraphics[width = 6cm]{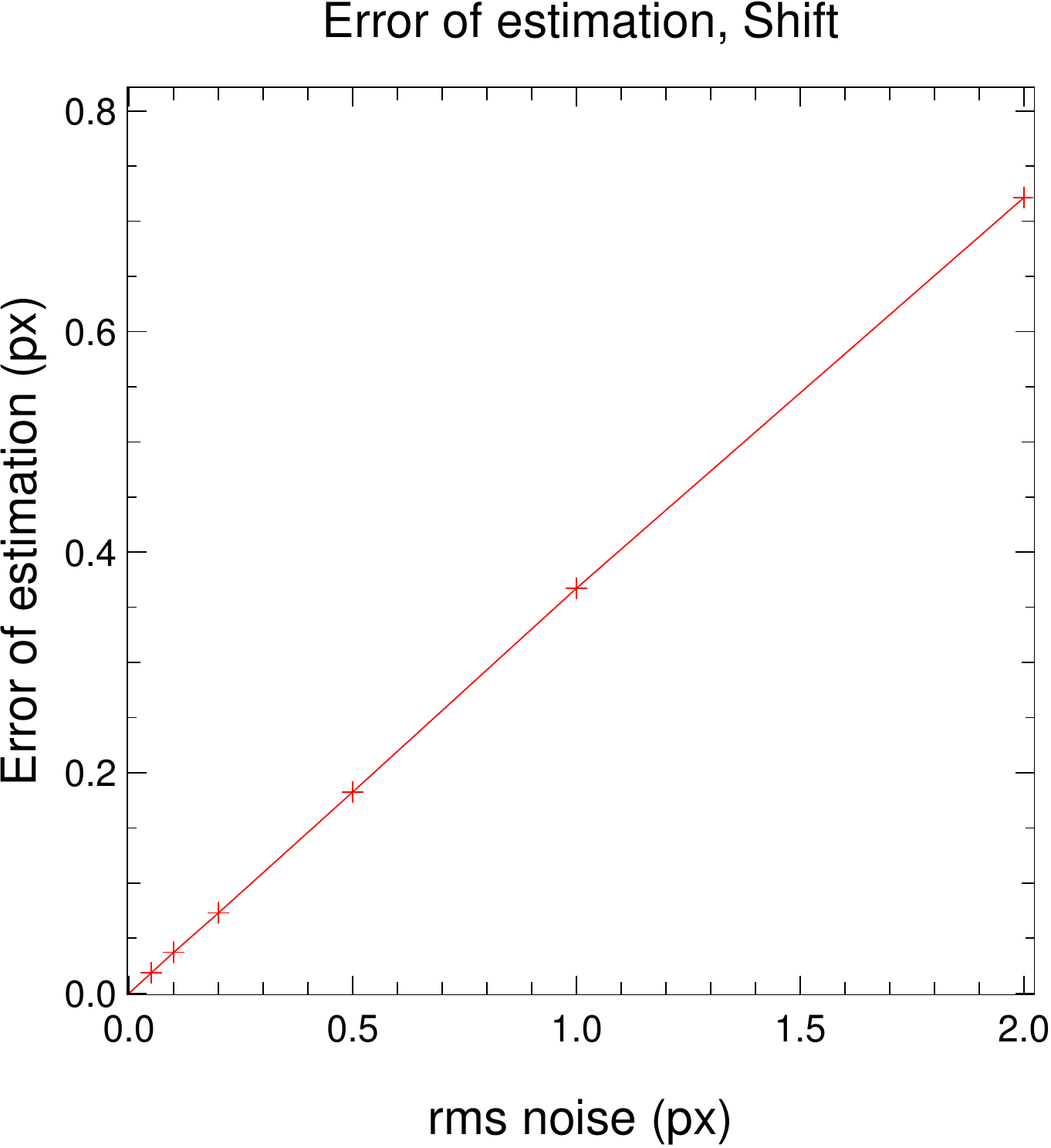}}   \hspace{1cm}	 \subfloat[]{\includegraphics[width = 6cm]{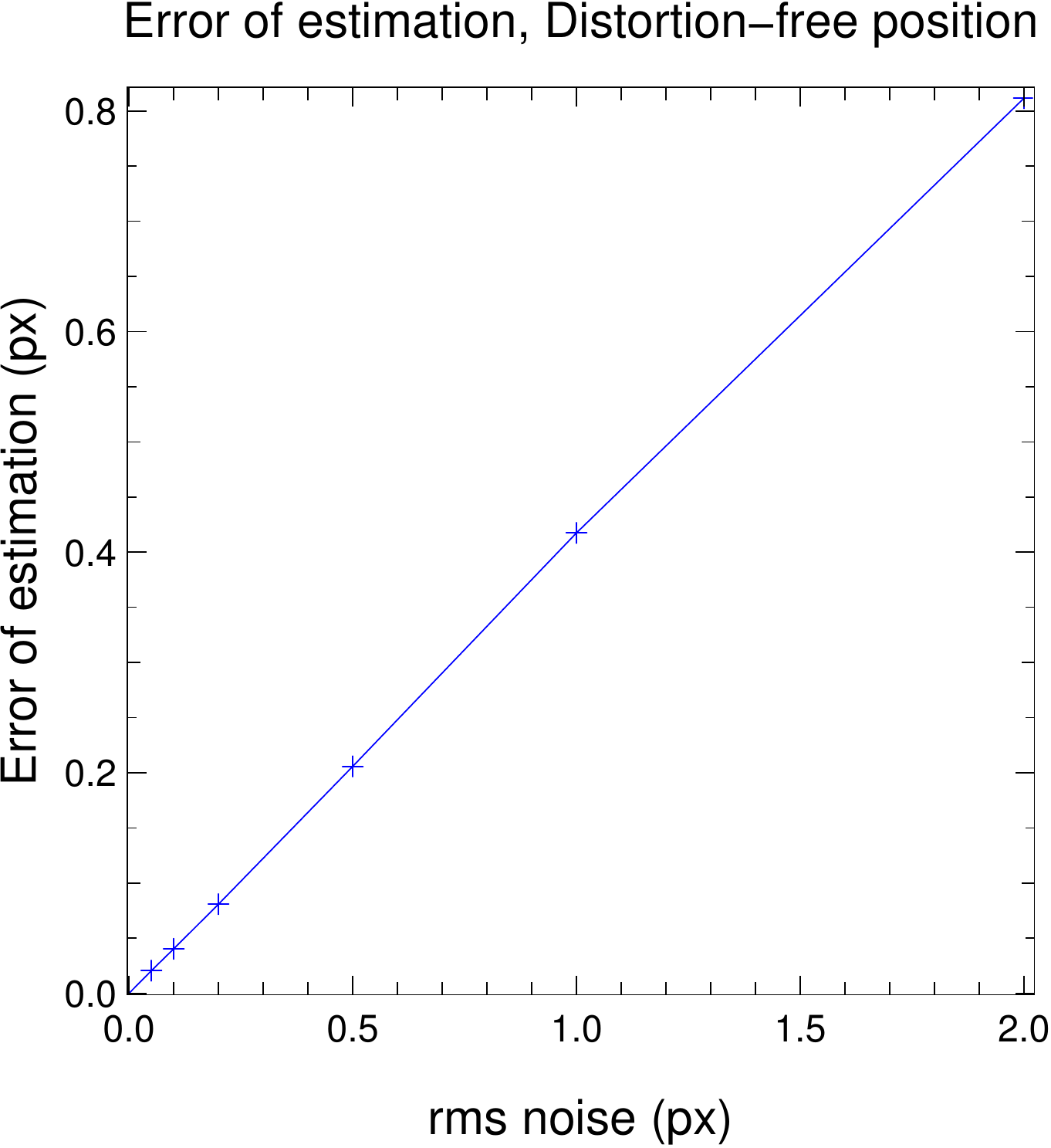}}
 \vspace{0.3cm}	
\caption{Error of estimation versus the noise standard deviation for simulations done with $N=3$, $Nim=10$ and $Nstar=20$. (a) Error on the first mode of distortion estimation: the shift (the error is averaged between the shift according to the $x$ axis and the shift according to the $y$ axis). (b) Error on the distortion-free position of the PSFs estimation. 
In both cases, the error represents the distance, in pixels, between the true value and the estimated value. It is quadratically averaged on the images and on the PSFs (respectively for the distortion coefficients and the distortion-free position of the PSFs) and on the 100 noise values. The errors are plotted versus the standard deviation of the added noise.}
\label{fig:noiseprop}
\end{center}
\end{figure}

We repeated this process for different sets of $[Nim, Nstar]$ and we found, as expected,  that the error on the distortion-free position of the PSFs is inversely proportional to $\sqrt{Nim}$ but independent on the number of stars, $Nstar$. On the contrary, the error on the distortion coefficients is inversely proportional to $\sqrt{Nstar}$ but independent on the number of frames, $Nim$. 

Finally, note that a having a $rms$ noise of 0.1 pixel on the measurement of a stars position on a typical set of GeMS/GSAOI data means dealing with a very faint object. For reference, a real data star fitted with a error of 0.1 pixel with a classical tool is shown in Fig.~\ref{fig:fit}. This means that even working exclusively with very faint objects, barley detectable, we still reach a very good precision on the estimation of the distortion coefficients and on the estimation of the distortion-free positions of the stars. The method then seems particularly well adapted for sparse field, allowing the use of every detectable star in the field.

\begin{figure*}[h!]
   \centering
 \includegraphics[scale=0.62]{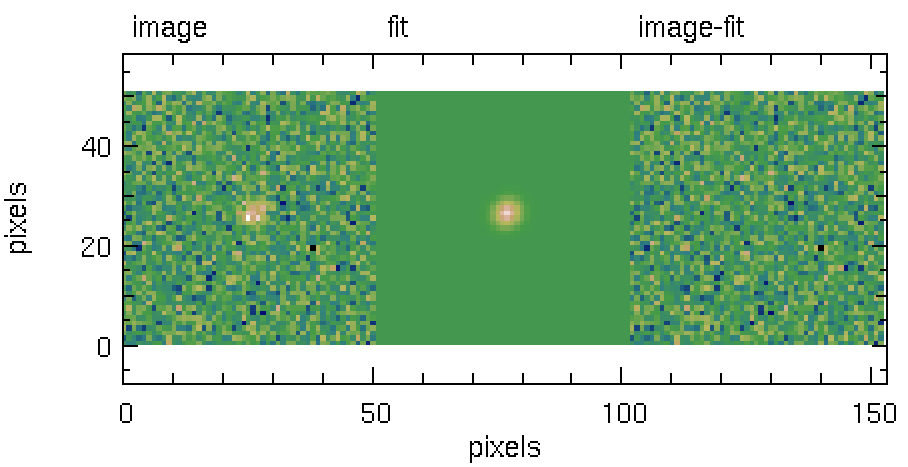}
  \vspace*{0.25cm}
   \caption{Example of a 0.1 pixel error fitted star from typical GeMS/GSAOI data. Example derived from the N159W data in \ks band.} 
             \label{fig:fit}%
\end{figure*}

\subsubsection{Correction of the distortion}

After estimating the distortion-free position of the PSFs and the distortion coefficients, we finally want to correct the different frames from their specific distortions. To do so, we use the estimated distortion coefficients $(A_i)_{\substack{1\leq i \leq Nim}}$ combined with a classical 2D interpolation function.
The correction is done independently frame by frame, therefore, the corrected position of one given PSF is variable from one frame to another. Thus, a good estimator of the correction efficiency is for each given PSF $j$, the standard deviation $\textit{std}_j$ of this PSF position through the stack. Besides, this criterion has the advantage that it is directly linked to the stacked image quality. 

\begin{equation*}
std_j = \sqrt{\frac{1}{Nim}\left(\sum\limits_{i=1}^{Nim} \left(X_{i,j,undist}-a\underset{i}vg~(X_{i,j,undist})\right)^2\right)}
\end{equation*}

With 
{$X_{i,j,undist}$, the corrected position of star $j$ in image $i$;

\begin{figure*}[h!]
   \centering
 \includegraphics[scale=0.5,trim={0 0cm 0cm 0cm}]{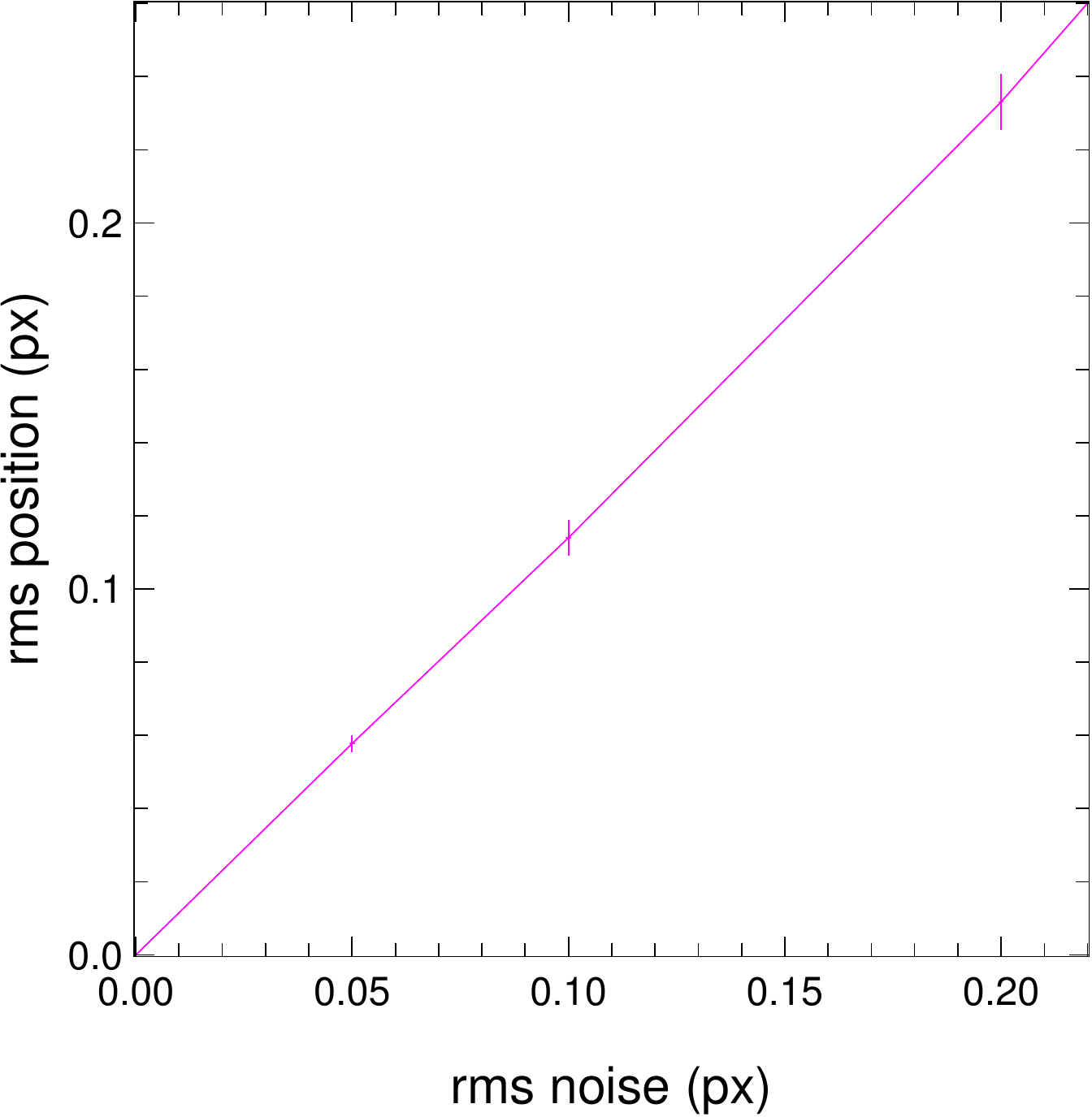}
  \vspace*{0.25cm}
   \caption{ Standard deviation of the PSF position between the different frames averaged quadratically on the 20 PSFs and the 100 draws, versus the \textit{rms} noise introduced on the measures. Both are in pixel.}
                 \label{fig:idl_y}%
\end{figure*}

Using the same set of data than described in Sect. \ref{sec:estimation}, we computed the standard deviation of each PSFs through the stack. Fig.~\ref{fig:idl_y} shows this standard  deviation averaged quadratically on the 20 PSFs and the 100 draws, versus the \textit{rms} noise introduced on the measurements. Both are in pixel. The typical error considering 0.1 pixel noise is 0.12 pixel i.e., about$\sim$2.4 mas.

\subsection{Application on GeMS data}

We present in this section, the first results obtained by applying the method of distortion correction describe above, on a set of ten random frames of N159W, the extra galactic star forming region observed with the Multi-Conjugated Adaptive Instrument GeMS/GSAOI (Gemini South Telescope) described in Sect. \ref{sec:sec1}.

As a first test, we selected the 46 brightest stars common to all frames and fitted them with a Levenberg–Marquardt algorithm in order to get their positions in each frame. We obtained a set of 46 x 10 stars positions, $(X^{data}_{i,j})_{\substack{1\leq i \leq 10\\1\leq j \leq 46}}$, fitted with a error lower than 0.1 pixel.
We estimated the distortion-free position of each of the 46 stars, $(X^{ref}_j)_{\substack{1\leq j \leq 46}}$ as well as the distortion coefficients associated to each frames $(A_i)_{\substack{1\leq i \leq 10}}$, and in a second time, we corrected each frame for its distortion as described in Sect. \ref{sec:simu}.

Fig. \ref{fig:xundist} (a) shows the estimated distortion-free positions (red crosses) as well as the standard deviation of the positions through the stack associated to each star (black circle).  The black circle in the bottom right corner correspond to a standard deviation of position through the stack of 0.1 pixel ($\sim$ 2 mas).
Fig. \ref{fig:xundist} (b) shows the final stacked image after correcting and combining the ten frames.

\begin{figure*}[h!]
   \centering
   \subfloat[]{\includegraphics[scale=0.41,trim={ 0cm 0cm 0cm 0cm}]{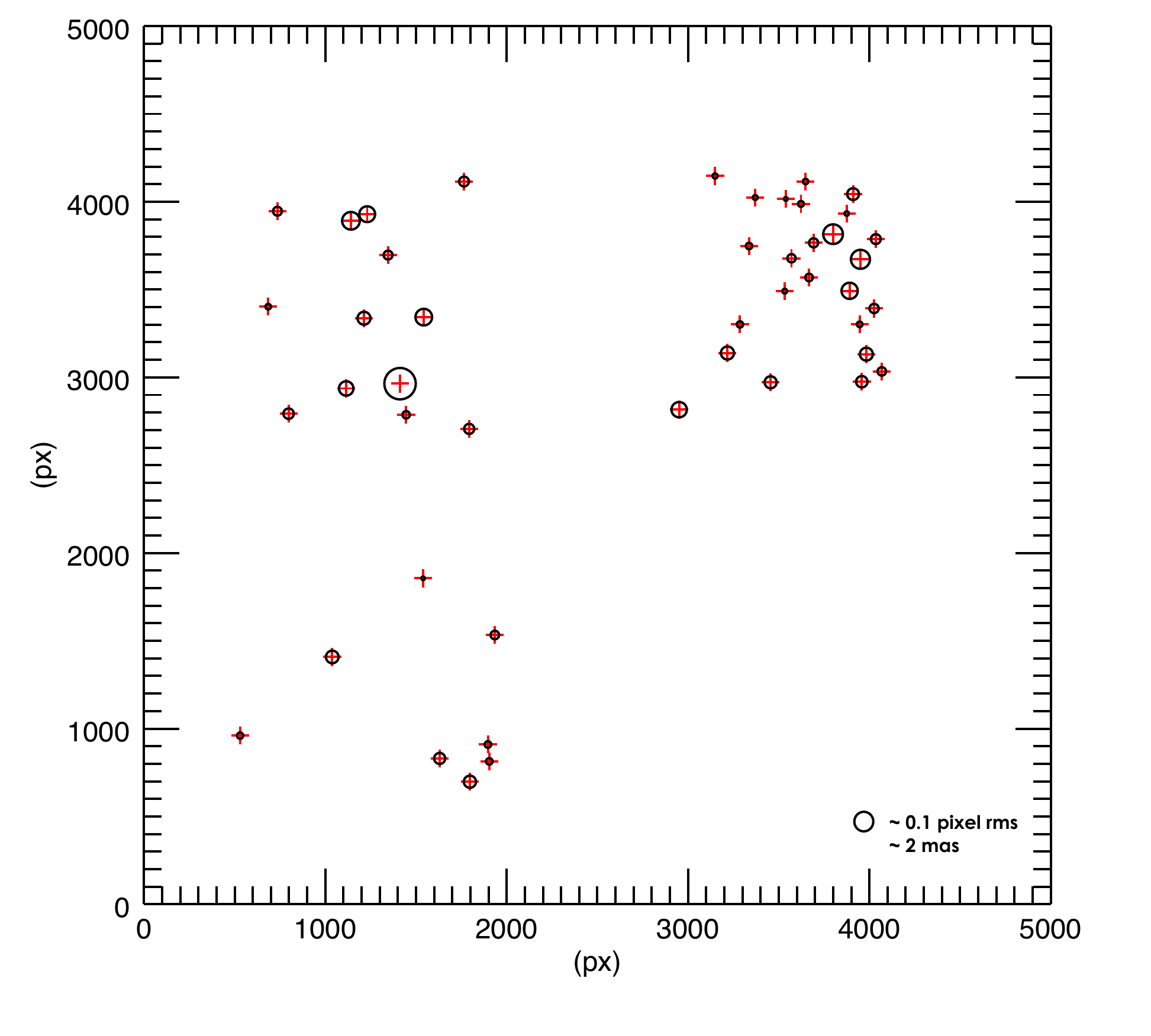}}   \hspace{1cm}\subfloat[]{\includegraphics[scale=0.5, trim={ 0cm -1cm 0cm 0cm}]{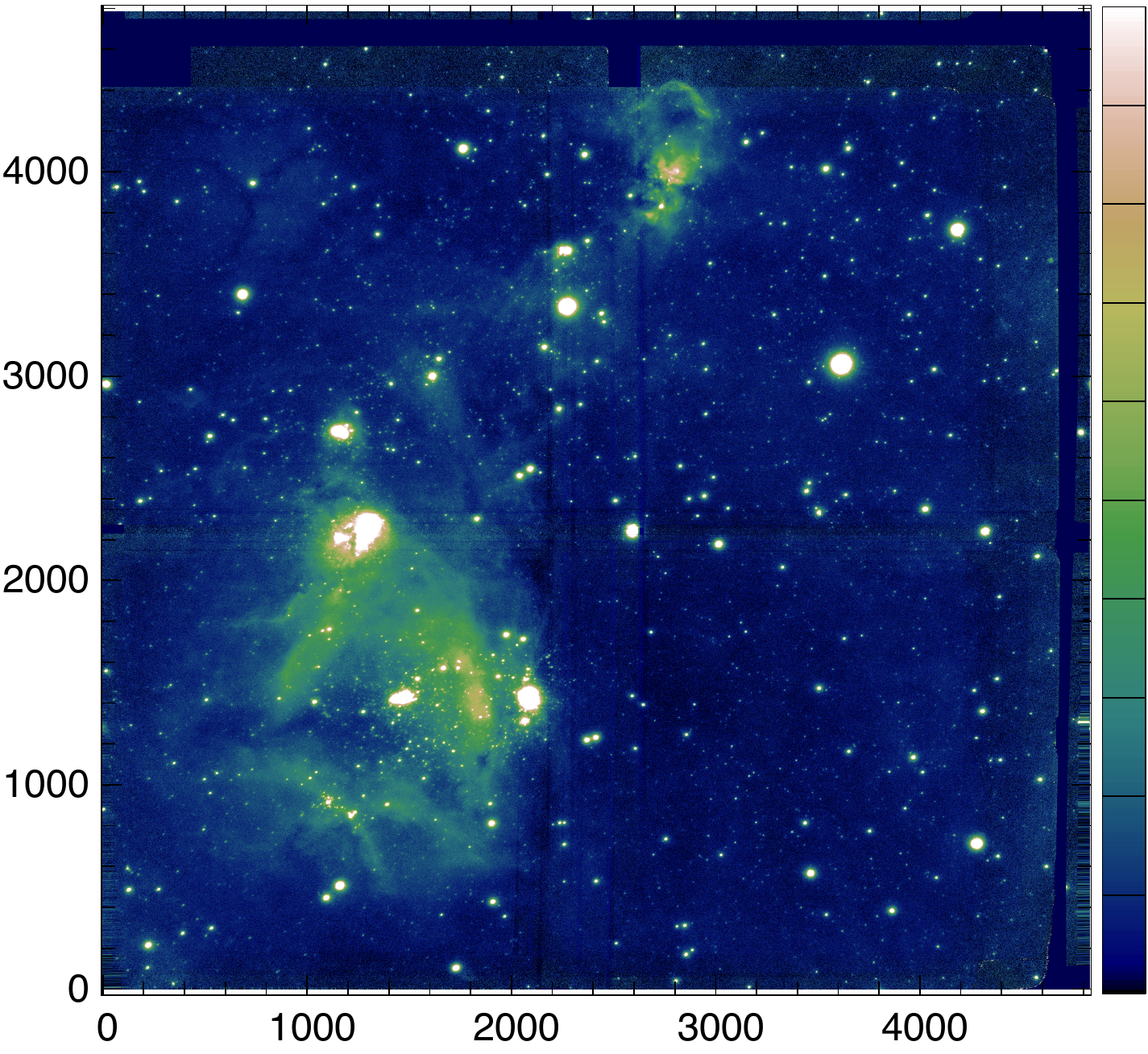}}
   \caption{ (a) Estimated distortion-free positions of the 46 stars used to correct the distortion (red crosses) and the standard deviation of the positions through the stack associated to each star after correction of the 10 frames (black circle). Black circle in the bottom right corner correspond to a standard deviation of position through the stack of 0.1 pixel ($\sim$ 2 mas). (b) Final stacked image combining 10 random frames of N159W.}
                 \label{fig:xundist}%
\end{figure*}	

As an indicator of the correction efficiency we also plotted the Full-Width-Half-Max (FWHM) distribution over the field for the stack of the 10 frames after correction of the distortion (Fig. \ref{fig:fwhm}). 

\begin{figure*}[h!]
   \centering
   \subfloat[]{\includegraphics[scale=0.5,trim={0 0 0 0cm 0},clip]{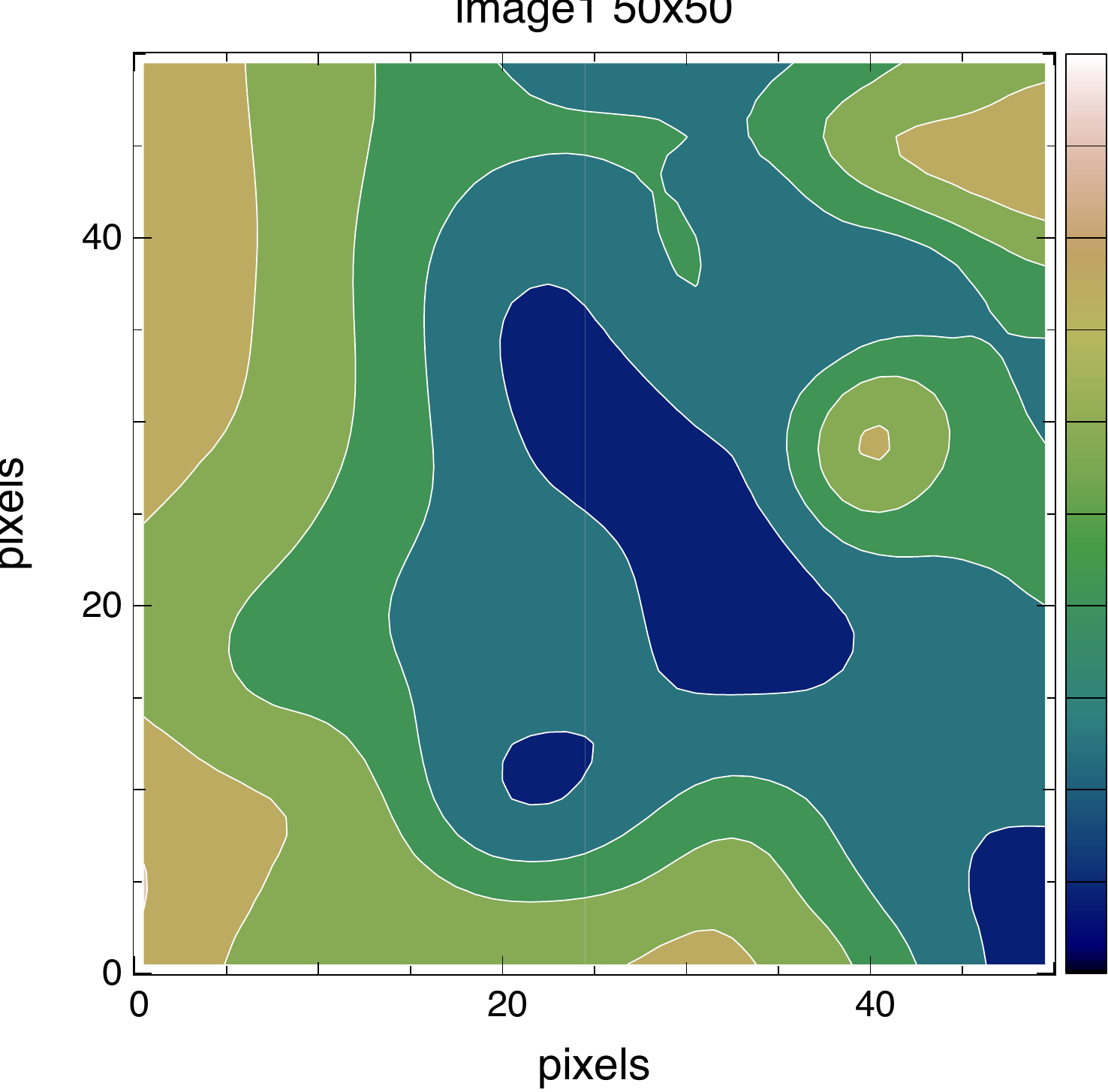}}
   \caption{Distribution of the Full-Width-Half-Max (FWHM) in the field of a ten frame stack of N159W with correction of distortion. Min is 80 mas (dark blue) and max is 106 mas (yellow).}
                 \label{fig:fwhm}%
\end{figure*}

The averaged FWHM in the field is about 160 mas without correction of distortion and about 90 mas with correction of distortion, which is similar to the FWHM of the individual frames ($\sim$90 mas).
This significant gain is very valuable for photometric studies. The next step is now to compare the estimated distortion-free positions, $(X^{ref}_j)_{\substack{1\leq j \leq 46}}$, with a distortion-free reference (as data obtained with the Hubble Space Telescope) to evaluate the astrometric performance of this method regarding real data. 


\section{CONCLUSION} 
In this paper we have presented deep, high angular resolution, near-infrared images of the N159W star forming region located in the Large Magellanic Cloud. The data were obtained with the Near-Infrared Wide Field Adaptive Optic instrument GeMS/GSAOI recently implemented on the Gemini South telescope. These images aim at exploring the stellar content of the cluster and the massive star formation history of the region by doing a photometric study.
Based on these real data, we evidence the limitations of the current reduction and analysis tools when it comes to Wide Field Adaptive Optics data. Indeed, despite the excellent performance of the adaptive optic correction, variable residuals are still limiting the quality of the image. Especially when combining multiple-frames, distortion effects consequently degrade the resolution and the sensitivity of the data. Thus, dedicated tools are needed to take into account these issues. 
In this article we presented a new method of correction of the distortion currently in development. This method is based on the least square minimization of a criterion, as previously done by D. Gratadour et al. (2005) \cite{Gratadour2005} for image re centering. We generalize here this method to any kind of distortion mode. The formalism as well as the implementation and performance results based on simulations as been presented.  We showed a strong robustness to the noise that allows the use of the faintest stars in the field to compute a reliable correction of the distortion. As the quality of the correction is proportional to the number of stars used to apply the correction, the robustness to the noise appears to be an essential parameter for sparse field  astrometry and photometry. Moreover, this method presents the advantage to estimate a "distortion-free" position of the stars in the sky, while existing tools base the correction on a non distortion-free reference. 
Finally, we have shown here a first application on real data. We estimated and corrected the distortion of ten random frames of the star forming region N159W. We found a typical standard deviation of position through the stack of 0.12 pixel $rms$ ($\sim$2.4 mas) which is consistent with simulations results. 
The perspectives for this work is now the optimization of several input parameters such as the number and the brightness of the stars used in the algorithm,  the consideration of their flux or magnitude using weighting coefficients, the number of degrees of freedom considered, and the choice of the distortion base. 
We also might consider to correct independently each GSAOI detector and, in a second time, put them together. 
Finally, in order to evaluate the astrometric reliability of this method regarding real data, it is now essential to compare the distortion-free positions estimated with some distortion-free reference e.g. data obtained with the Hubble Space Telescope. 
\acknowledgments

Based on observations obtained at the Gemini Observatory, which is operated by the Association of Universities for Research in Astronomy, Inc., under a cooperative agreement with the NSF on behalf of the Gemini partnership: the National Science Foundation (United States), the National Research Council
(Canada), CONICYT (Chile), the Australian Research Council (Australia), Minist\'erio da Ci$\rm{\hat{e}}$ncia, Tecnologia
e Inova{\c{c}}$\rm{\tilde{a}}$o (Brazil) and Ministerio de Ciencia, Tecnologiaa e Innovaci\'on Productiva (Argentine).\\ 
The work was partly funded by the European Commission under FP7 Grant Agreement No. 312430 Optical
Infrared Coordination Network for Astronomy and by the Office National d’Etudes et de Recherches Arospatiales
(ONERA).\\
B. Neichel and A. Bernard acknowledge the financial support from the French ANR program WASABI to carry out this work.
\bibliography{library_ab_spie.bib}
\bibliographystyle{spiebib} 

\end{document}